\documentclass[runningheads]{llncs}
\usepackage[T1]{fontenc}
\usepackage{amsmath}
\usepackage{adjustbox}
\usepackage{tabularx}
\usepackage{graphicx}
\begin{document}
\title{M-Vec: Matryoshka Speaker Embeddings with Flexible Dimensions}
%
%
\author{Shuai Wang\inst{1,2}\orcidID{0000-0003-1523-9631}
\and Pengcheng Zhu\inst{3}\orcidID{0009-0000-8026-9824}$\dagger$ 
\and Haizhou Li\inst{2,1}\orcidID{0000-0001-9158-9401}}
\authorrunning{S. Wang, P. Zhu and H. Li}
%
\institute{Shenzhen Research Institute of Big Data \and
School of Data Science, Chinese University of Hong Kong (Shenzhen) 
\and
Fuxi AI Lab, NetEase Inc., Hangzhou, China
\\
\email{\{wangshuai,haizhouli\}@cuhk.edu.cn, zhupengcheng@corp.netease.com}}

\maketitle              
\begin{abstract}
Fixed-dimensional speaker embeddings have become the dominant approach in speaker modeling, typically spanning hundreds to thousands of dimensions. These dimensions are hyperparameters that are not specifically picked, nor are they hierarchically ordered in terms of importance. In large-scale speaker representation databases, reducing the dimensionality of embeddings can significantly lower storage and computational costs. However, directly training low-dimensional representations often yields suboptimal performance. In this paper, we introduce the Matryoshka speaker embedding, a method that allows dynamic extraction of sub-dimensions from the embedding while maintaining performance. Our approach is validated on the VoxCeleb dataset, demonstrating that it can achieve extremely low-dimensional embeddings, such as 8 dimensions, while preserving high speaker verification performance.

\end{abstract}
\section{Introduction}

\footnotetext{$\dagger$ Correspondence to: Pengcheng Zhu (zhupengcheng@corp.netease.com)}

\subsection{Speaker Modeling for Human-Computer Interaction}
Speech is one of the primary modalities for human-computer interaction and closely aligns with natural human-to-human communication methods. Voice interaction is suitable for various scenarios, including those where hands-free operation or visual impairment is necessary. Among the attributes of speech signals, the identity of the speaker is one of the most critical pieces of information. Preemptively identifying the speaker is a crucial step in delivering customized and personalized services. Speaker recognition in speech technology significantly enhances the intelligence, personalization, and security of human-computer interaction. Speaker recognition encompasses two tasks: speaker identification and speaker verification. Speaker identification selects the matching speaker from a list of candidates, while speaker verification determines if the registered and test voices are from the same person.

\subsection{Background on speaker embedding learning}
Currently, speaker representation is primarily expressed using a fixed-dimensional embedding~\cite{speaker_overview}. Before the advent of deep learning, this embedding was typically learned using factor analysis-based methods, with the corresponding representation known as the i-vector~\cite{ivector}. In the era of deep learning, neural networks are employed to compress the speech signal and extract speaker information. Generally, the entire neural network is trained with the optimization objective of speaker classification, ensuring that the extracted representation possesses sufficient speaker discriminability. Notable neural network frameworks include Time Delay Neural Networs (TDNN)~\cite{tdnn,xvec}, ResNet~\cite{resnet,rvec}, and ECAPA-TDNN~\cite{ecapa}. 

\subsection{Extremely Low-Dimensional Embeddings }
During large-scale database searches, the dimensionality of the representation directly influences the storage costs and is closely related to the efficiency of the search process. Consequently, many researchers are exploring ways to model information using the most compact vectors possible.

On the other hand, traditional i-vectors and subsequent deep speaker embeddings, such as i-vectors~\cite{ivector} with 400 or 600 dimensions, x-vectors~\cite{xvec} based on TDNN with 512 dimensions, or r-vectors~\cite{rvec} based on ResNet with 256 dimensions, are often set based on empirical values. Each dimension does not represent a specific meaning, and there is no distinction in importance, making it challenging to obtain low-dimensional representations by directly filtering the embedding dimensions. Li et al.~\cite{liimportant} has trained an additional autoencoder for importance rearrangement to transform features, but this approach requires multiple stages in the extraction of speaker embeddings.

\subsection{Contributions}
In this paper, we introduce the concept of Matryoshka representation learning and apply it within the current mainstream AAM-loss learning framework. The main contribution can be sumarized as following,
\begin{enumerate}
    \item We are the first to explore variable-dimensional speaker representations, where representations of different dimensions can be adapted to various related tasks.
    \item We propose the training method Matryoshka Representation Learning (MRL), which enables simultaneous speaker discriminative training across multiple dimensions.
    \item Our proposed structure significantly enhances the modeling capabilities of representations, even in extremely low-dimensional cases (e.g., 4-dimensional or 8-dimensional).
\end{enumerate}

\section{Matryoshka Embedding Learning}
Traditional approaches typically employ fixed, full-sized embeddings for all tasks, disregarding the variations in resource constraints and requirements across different applications. This practice can result in computational inefficiency and poor scalability when dealing with the diverse resource availabilities in downstream scenarios. To address this challenge, Matryoshka Representation Learning (MRL)~\cite{mrl} introduces an innovative methodology: the concurrent training of multiple embeddings with nesting dimensions, thereby achieving scalable embedding sizes.

Inspired by Matryoshka dolls, where smaller dolls nest within larger ones, MRL similarly nests smaller embeddings within larger ones, enabling a single model to generate embeddings of varying sizes. This approach offers flexibility in computational resource utilization and adaptability to diverse application requirements.

\subsection{Matryoshka Representation}
The Matryoshka representation learning loss function for the first $m$ dimensions is as follows:

\begin{equation}
   \mathcal{L}_m =  \min_{\{\mathbf{W}(m)\}_{m \in M}, \theta_F} \sum_{m \in M} c_m \cdot \mathcal{L}(\mathbf{W}(m) \cdot F(x; \theta_F)_{1:m}; y)
\end{equation}

In this formula:
\begin{itemize}
    \item $\mathbf{W}(m) \cdot F(x; \theta_F)_{1:m}$ represents feeding the first \( m \) dimensions of the embedding vector $F(x; \theta_F)_{1:m}$ to the linear classifier $\mathbf{W}(m)$ for classification.
    \item $F$ is a feature extractor to transform input $x$ to the embedding space.
    \item $c_m$ is the loss weight for the first $m$ dimensions.
    \item $\mathcal{L}$ is the multi-class cross-entropy loss function used to measure the difference between the classifier's output and the true label $y$.
    \item By minimizing this loss function, we can simultaneously optimize the neural network parameters $\theta_F$ and the weights of the linear classifiers $\{\mathbf{W}(m)\}_{m \in M}$ to ensure that the embedding vectors at different granularities have strong discriminative power.
\end{itemize}

\subsection{MRL for Speaker Embedding Learning}
In this work, we introduce the concept of Matryoshka Representation Learning (MRL) into the speaker embedding learning framework.

\begin{figure}[ht]
    \centering
    \includegraphics[width=0.8\textwidth]{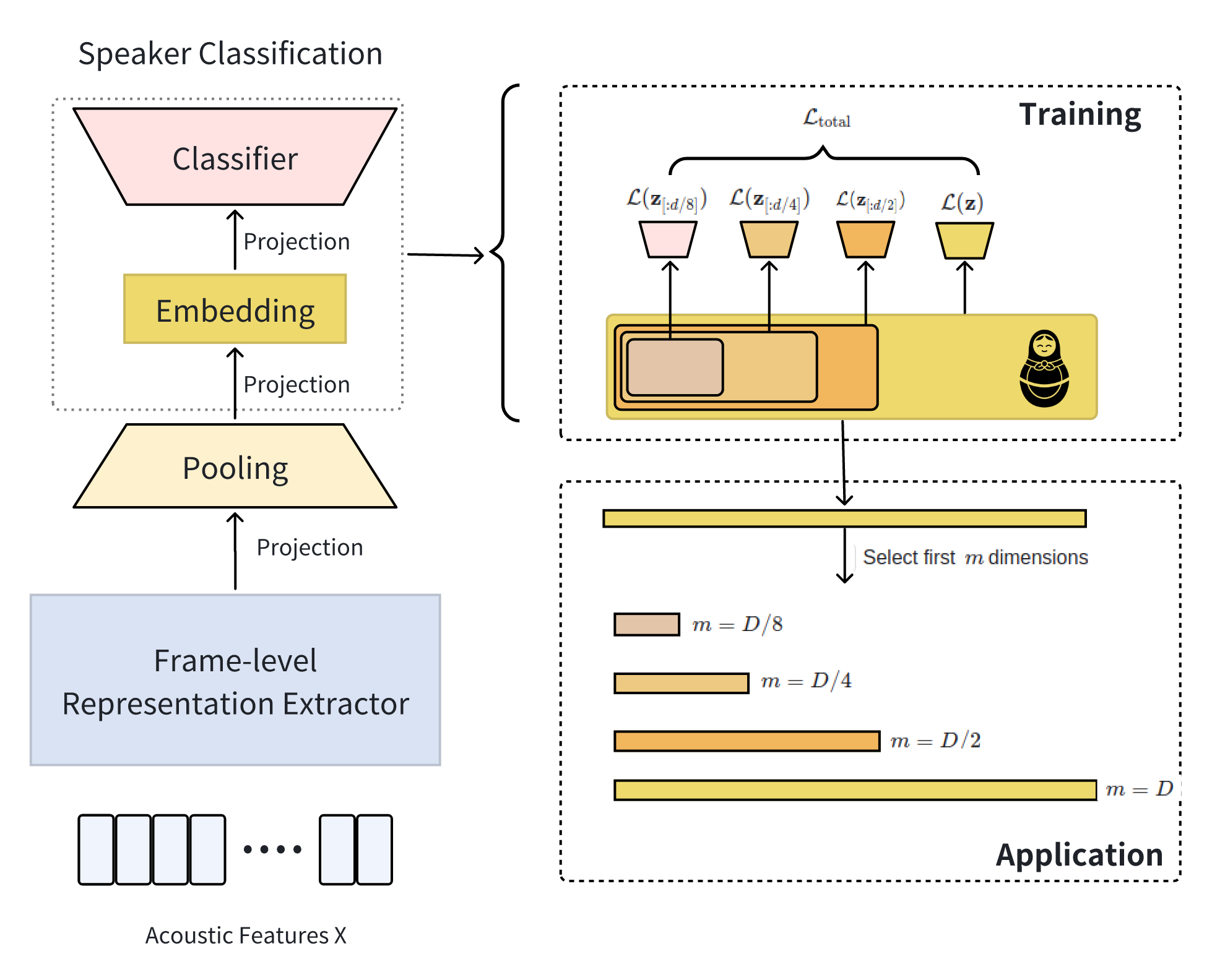}
    \caption{Matryoshca Speaker Embedding Learning, we use 3 sub-dimensional embeddings as a illustration}
    \label{fig:mrl}
\end{figure}

As shown in Figure~\ref{fig:mrl}, the standard speaker representation extractor is illustrated in the left half of the figure. Frame-level features are extracted from wave files and encoded into frame-level deep features by a frame-level representation extractor, which maintains the temporal resolution. Next, a pooling layer is employed to aggregate frame-level features into segment-level features, which are subsequently projected to lower-dimensional speaker embeddings. The entire network is optimized for speaker classification loss, typically using functions from the softmax family, such as the original softmax or margin-based variants like AAM-softmax~\cite{aamsoftmax}\cite{margin}. In this paper, we utilize AAM-Softmax.

In the improved system, by introducing the concept of MRL, we explicitly optimize the\emph{ nested sub-dimensions} of the entire speaker embedding. This approach allows for the flexible extraction of the first few dimensions as needed while maintaining performance. Specifically, we integrate the commonly employed AAM-Softmax loss function with the Matryoshka Representation Learning (MRL) framework. This novel loss function is then utilized to learn Matryoshka speaker embeddings.

AAM-Softmax (Additive Angular Margin Softmax) improves upon the standard Softmax loss by introducing an angular margin to enhance inter-class separation.
The AAM-Softmax loss function with the embedding vector \( \mathbf{e} \) and the margin \( k \) can be expressed as:

\begin{equation}
   \mathcal{L}_{\text{AAM}} = -\frac{1}{N} \sum_{i=1}^N \log \frac{e^{s \left( \frac{\mathbf{W}_{y_i}^T \mathbf{e}_i}{\|\mathbf{W}_{y_i}\| \|\mathbf{e}_i\|} + k \right)}}{e^{s \left( \frac{\mathbf{W}_{y_i}^T \mathbf{e}_i}{\|\mathbf{W}_{y_i}\| \|\mathbf{e}_i\|} + k \right)} + \sum_{j \neq y_i} e^{s \frac{\mathbf{W}_j^T \mathbf{e}_i}{\|\mathbf{W}_j\| \|\mathbf{e}_i\|}}}
\end{equation}

In this formula:
\begin{itemize}
    \item \( \mathbf{W}_{y_i} \) is the weight vector for the target class \( y_i \).
    \item \( \mathbf{W}_j \) are the weight vectors for the non-target classes.
    \item \( \mathbf{e}_i \) is the embedding vector for the \( i \)-th sample.
    \item \( s \) is the scaling factor.
    \item \( k \) is the additive angular margin.
\end{itemize}

The MRL augmented version of AAM-softmax can be expressed as,

\begin{equation}
    \mathcal{L}_m = \min_{\{\mathbf{W}(m)\}_{m \in M}, \theta_F} \sum_{m \in M} c_m \cdot \mathcal{L}_{\text{AAM}}\left(\mathbf{W}(m), \{\mathbf{e}_{i,1:m}\}_{i=1}^N, s, k\right)
\end{equation}

Where:

\begin{itemize}
    \item \( \mathcal{L}_{\text{AAM}}\left(\mathbf{W}(m), \{\mathbf{e}_{i,1:m}\}_{i=1}^N, s, k\right) \) is the AAM-Softmax loss function, which measures the difference between the classifier's output and the true label \( y \). Specifically, the AAM-Softmax loss function is given by:
    \begin{equation}
    \mathcal{L}_{\text{AAM}} = -\frac{1}{N} \sum_{i=1}^N \log \frac{e^{s \left( \frac{\mathbf{W}_{y_i}(m)^T \mathbf{e}_{i,1:m}}{\|\mathbf{W}_{y_i}(m)\| \|\mathbf{e}_{i,1:m}\|} + k \right)}}{e^{s \left( \frac{\mathbf{W}_{y_i}(m)^T \mathbf{e}_{i,1:m}}{\|\mathbf{W}_{y_i}(m)\| \|\mathbf{e}_{i,1:m}\|} + k \right)} + \sum_{j \neq y_i} e^{s \frac{\mathbf{W}_j(m)^T \mathbf{e}_{i,1:m}}{\|\mathbf{W}_j(m)\| \|\mathbf{e}_{i,1:m}\|}}}
    \end{equation}
    \item \( \mathbf{W}(m) \cdot \mathbf{e}_{i,1:m} \) represents feeding the first \( m \) dimensions of the embedding vector \( \mathbf{e}_{i,1:m} \) to the linear classifier \( \mathbf{W}(m) \) for classification.
    \item $c_m$ is the loss weight for the first $m$ dimensions.
    \item \( \theta_{y_i}^m \) is the angle between the embedding \( \mathbf{e}_{i,1:m} \) and the weight vector \( \mathbf{W}_{y_i}(m) \) for the target class.
    \item \( \theta_j^m \) are the angles between the embedding \( \mathbf{e}_{i,1:m} \) and the weight vectors \( \mathbf{W}_j(m) \) for all other classes.
\end{itemize}

\section{Experiments}
\subsection{Dataset}
The VoxCeleb dataset, introduced by Oxford University, has become one of the most extensively used text-independent speaker recognition datasets in the field. In this work, we adopt the ``dev'' partition of VoxCeleb2 as the training set and whole VoxCeleb1 as the test set. Equal error rate (EER) is used to evaluate the performance on the speaker verification task.

\subsection{Experimental Setups}
All experiments in this study were conducted using the \texttt{wespeaker} toolkit~\cite{wespeaker}, adhering to the data preparation protocols outlined in its VoxCeleb recipe. Audio samples from the MUSAN dataset~\cite{musan} served as additive noise sources, while simulated room impulse responses (RIRs)\footnote{\url{https://www.openslr.org/28}} were utilized to introduce reverberation effects. For each training set utterance, we applied either noise or reverberation augmentation (but not both concurrently) with a probability of 0.6. Additionally, speed perturbation was performed by altering the speed of an utterance to 0.9x or 1.1x, with the resultant audio being treated as originating from new speakers due to the pitch shift caused by the augmentation.

Following data preparation, two baseline systems, TDNN and ResNet34 were implemented. Detailed optimization strategies for these systems can be found in the respective recipes provided by WeSpeaker\footnote{\url{https://github.com/wenet-e2e/wespeaker}}.

For all experiments, we set $c_m$ equally to 1, $M=\{8,16,32,64,128,256\}$

\subsection{Results and Analysis}
The experimental results can be found in Table~\ref{table:performance_comparison}. As shown, the ResNet34 system significantly outperforms the TDNN system in full dimension scenarios. However, both systems exhibit similar performance when evaluated under extremely low-dimensional settings. We used the more powerful ResNet34 as main system for validate our algorithm, with the corresponding results represented by the ``ResNet34-MRL''.

\begin{table}[ht]
\centering
\begin{adjustbox}{max width=1.5\textwidth}
\begin{tabularx}{\textwidth}{l|X|X|X|X|X|X}
\hline
Model         & 8      & 16     & 32     & 64     & 128    & 256    \\ \hline
TDNN          & 18.99 & 10.59 & 7.526  & 4.971  & 3.068  & 2.579  \\ \hline
ResNet34      & 18.78 & 10.41 & 5.733  & 2.499  & 1.420  & \textbf{1.124}  \\ \hline
\textbf{ResNet34-MRL } & \textbf{4.941}  & \textbf{2.605}  & \textbf{1.574}  & \textbf{1.313}  & \textbf{1.154}  & 1.153  \\ \hline
\end{tabularx}
\end{adjustbox}
\caption{Performance Comparison Across Different Dimensions}
\label{table:performance_comparison}
\end{table}

To more intuitively demonstrate the performance of different systems at various embedding dimensions, we visualize the results from Table 1 in Figure~\ref{fig:performance}. It is evident that as the embedding dimension decreases, our proposed system, ResNet34-MRL, exhibits a significantly smoother performance curve.


\subsubsection{Comparison of Embeddings with Different Dimensions}
\begin{figure}[ht]
    \centering
    \includegraphics[width=0.95\textwidth]{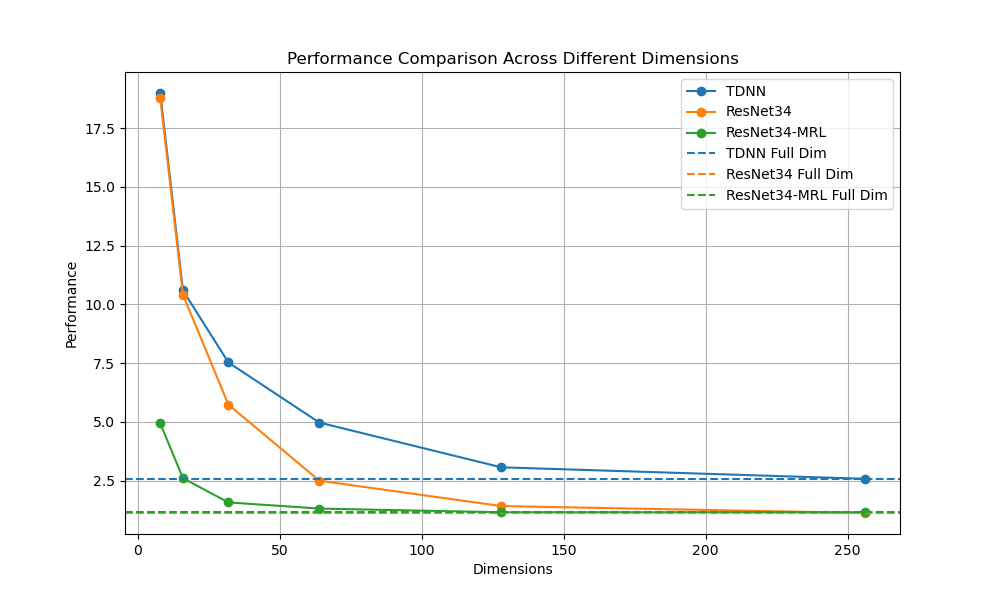}
    \caption{Performance comparison of different systems using different dimensions}
    \label{fig:performance}
\end{figure}

To more intuitively demonstrate the ability of the MRL method to preserve performance across different dimensions, we have visualized the results from Table~\ref{table:performance_comparison} in Figure~\ref{fig:performance}. It is evident that, across all dimensions except the full-version, our proposed ResNet34-MRL method consistently outperforms the ResNet34 system. Particularly in low-dimensional scenarios, we do not observe the dramatic performance degradation seen in the baseline systems, further proving the effectiveness of our approach. For the full dimension of 256, we observe a slight performance degradation, with the EER increased from $1.124\%$ to $1.153\%$ 

\subsubsection{Extremely Low Dimensional Embeddings}

We would like to emphasize the relatively strong performance of dimensions lower than 32, highlighting the remarkable effectiveness of the MRL strategy in extremely low-dimensional settings. The 16-dimensional ResNet34-MRL embeddings achieve results comparable to the TDNN system using the full 256 dimensions, with an EER of $2.605\%$.
Moreover, even in the extreme case of 8 dimensions, it achieves a relatively good performance with an EER of 4.941\%.
\subsection{Analysis on the Storage and Retivial Time} 

In the table below, we quantify the storage space usage and retrieval speed for different embedding dimensions.  We utilized the Faiss library, a mature and widely adopted solution in many commercial systems, to demonstrate this point. The CPU version of Faiss~\footnote{\url{https://github.com/facebookresearch/faiss}} is used as the similarity-based search solution. The similarity measurement used is `L2'\footnote{We first normalize all speaker embeddings and then do the L2 distance computation, which is equivalent to the cosine similarity. }
\footnote{\url{https://github.com/facebookresearch/faiss/wiki/MetricType-and-distances}}, and each time we retrieve the top 10 most similar entries for the given reference, from a database with \emph{10 million }candidate embeddings. The CPU we used for running the Faiss tests is an `\texttt{Intel(R) Xeon(R) Silver 4210R CPU @ 2.40GHz}'.

\begin{table}[h!]
\centering
\begin{adjustbox}{max width=\textwidth}
\begin{tabular}{c|c|c|c|c}
\hline
\textbf{ Dimension} & \textbf{Storage (MB)} & \textbf{Retrieval Time (ms)} & \textbf{$\Delta$ Storage (\%)}$\downarrow$ & \textbf{$\Delta$ Retrieval Time (\%)}$\downarrow$ \\
\hline
256 & 9765.62 & 759.31 & 0.00 & 0.00 \\
\hline
128 & 4882.81 & 377.42 & 50.00 & 50.29 \\
\hline
64 & 2441.41 & 194.01 & 75.00 & 74.45 \\
\hline
32 & 1220.70 & 115.18 & 87.50 & 84.83 \\
\hline
16 & 610.35 & 76.86 & 93.75 & 89.88 \\
\hline
8 & 305.18 & 47.46 & 96.88 & 93.75 \\
\hline
\end{tabular}
\end{adjustbox}
\caption{Comparison of Storage and Retrieval Time under Different Dimensions}
\label{table:performance_comparison}
\end{table}
By employing the Matryoshka speaker embedding strategy and selecting sub-dimensions, we can significantly reduce the demands on storage and computational resources. This approach not only helps lower storage costs but also enhances retrieval efficiency, making it highly valuable for constructing and querying large-scale speaker representation databases. The storage requirements are linearly related to the embedding dimensions, and retrieval efficiency based on similarity comparisons follows a similar linear relationship. 

\section{Conclusion}
In this paper, we propose the Matryoshka speaker embedding learning strategy, which allows users to flexibly customize the embedding dimensions during inference without the need to retrain the model. We also ensure the discriminability of the representations at extremely low dimensions. On the VoxCeleb1 test set, using only an 8-dimensional embedding, we achieve an EER of 4.9\%, and with a 16-dimensional embedding, we achieve an EER of 2.6\%. This strategy can be extended to any speaker encoder. The extremely low-dimensional representations learned through our method can significantly reduce storage requirements and retrieval times.

\newpage

\begin{credits}
\subsubsection{\ackname} This work is supported by Internal Project of Shenzhen Research Institute of Big Data under grant No. T00120220002 and No.J00220230014; and CCF-NetEase ThunderFire Innovation Research Funding (No. CCF-Netease 202302).
\end{credits}
%
%
%

\begin{thebibliography}{8}

\bibitem{mrl}
Kusupati A, Bhatt G, Rege A, et al. Matryoshka representation learning[J]. Advances in Neural Information Processing Systems, 2022, 35: 30233-30249.

\bibitem{wespeaker}
Wang H, Liang C, Wang S, et al. Wespeaker: A research and production oriented speaker embedding learning toolkit[C]//ICASSP 2023-2023 IEEE International Conference on Acoustics, Speech and Signal Processing (ICASSP). IEEE, 2023: 1-5.

\bibitem{wespeakerj}
Wang S, Chen Z, Han B, et al. Advancing speaker embedding learning: Wespeaker toolkit for research and production[J]. Speech Communication, 2024: 103104.

\bibitem{musan}
Snyder D, Chen G, Povey D. Musan: A music, speech, and noise corpus[J]. arXiv preprint arXiv:1510.08484, 2015.

\bibitem{ivector}
Dehak N, Kenny P J, Dehak R, et al. Front-end factor analysis for speaker verification[J]. IEEE Transactions on Audio, Speech, and Language Processing, 2010, 19(4): 788-798.

\bibitem{resnet}
He K, Zhang X, Ren S, et al. Deep residual learning for image recognition[C]//Proceedings of the IEEE conference on computer vision and pattern recognition. 2016: 770-778.

\bibitem{tdnn}
Peddinti V, Povey D, Khudanpur S. A time delay neural network architecture for efficient modeling of long temporal contexts[C]//Interspeech. 2015: 3214-3218.

\bibitem{rvec}
Zeinali H, Wang S, Silnova A, et al. But system description to voxceleb speaker recognition challenge 2019[J]. arXiv preprint arXiv:1910.12592, 2019.

\bibitem{xvec}
Snyder D, Garcia-Romero D, Sell G, et al. X-vectors: Robust dnn embeddings for speaker recognition[C]//2018 IEEE international conference on acoustics, speech and signal processing (ICASSP). IEEE, 2018: 5329-5333.

\bibitem{speaker_overview}
Wang S, Chen Z, Lee K A, et al. Overview of Speaker Modeling and Its Applications: From the Lens of Deep Speaker Representation Learning[J]. arXiv preprint arXiv:2407.15188, 2024.

\bibitem{ecapa}
Desplanques B, Thienpondt J, Demuynck K. Ecapa-tdnn: Emphasized channel attention, propagation and aggregation in tdnn based speaker verification[J]. arXiv preprint arXiv:2005.07143, 2020.

\bibitem{liimportant}
Li L, Xing C, Wang D, et al. Binary speaker embedding[C]//2016 10th International Symposium on Chinese Spoken Language Processing (ISCSLP). IEEE, 2016: 1-4.

\bibitem{aamsoftmax}
Deng J, Guo J, Xue N, et al. Arcface: Additive angular margin loss for deep face recognition[C]//Proceedings of the IEEE/CVF conference on computer vision and pattern recognition. 2019: 4690-4699.

\bibitem{margin}
Xiang X, Wang S, Huang H, et al. Margin matters: Towards more discriminative deep neural network embeddings for speaker recognition[C]//2019 Asia-Pacific Signal and Information Processing Association Annual Summit and Conference (APSIPA ASC). IEEE, 2019: 1652-1656.

\end{thebibliography}
%

\end{document}